\documentclass{article}

\usepackage{PRIMEarxiv}

\usepackage[utf8]{inputenc}
\usepackage[T1]{fontenc}

\usepackage{hyperref}
\usepackage{url}
\usepackage{booktabs}
\usepackage{amsmath,amssymb,amsfonts}
\usepackage{nicefrac}
\usepackage{microtype}
\usepackage{lipsum}
\usepackage{fancyhdr}
\usepackage{graphicx}
\graphicspath{{media/}}
\usepackage{cite}
\usepackage[most]{tcolorbox}
\usepackage{multirow}
\usepackage{multicol}
\usepackage{enumitem}
\usepackage{fontawesome5}
\usepackage{array}
\usepackage{algorithmic}
\usepackage{textcomp}
\usepackage{comment}
\usepackage{xspace}
\usepackage{xcolor}

\usepackage{tabularx}
\usepackage{arydshln}

\newcolumntype{Y}{>{\centering\arraybackslash}X}
\newcommand{\utech}{shadow IT\xspace}

\newcommand{\qoute}[1]{\textit{#1}}
\renewcommand{\sp}[0]{security and privacy\xspace}
\newcommand{\etal}{\textit{et al.}}

\newcounter{boxcount}
\newcommand{\SummaryBox}[1]{
\begin{tcolorbox}[colback=yellow!10, colframe=black!75!white, left=0.3em, right=0.3em, top=0.3em, bottom=0.3em,
]
\faIcon{edit} #1

\end{tcolorbox}
\stepcounter{boxcount}
}

\title{Investigating Shadow IT Use and Adoption in \\
Educational Institutions and their Privacy Implications
\thanks{This material is based upon work supported by the National Science Foundation under Grants 2544663 and 2350036.}
}

\author{
\textbf{Easton Kelso}\\
\texttt{eakelso@asu.edu}\\
Arizona State University
\and
\textbf{Ananta Soneji}\\
\texttt{asoneji@asu.edu}\\
Arizona State University
\and 
\textbf{Yan Soshitaishvili}\\
\texttt{yans@asu.edu}\\
Arizona State University
\and\\
\textbf{Sazzadur Rahaman}\\
\texttt{sazz@cs.arizona.edu}\\
University of Arizona\\
\and\\
\textbf{Rakibul Hasan}\\
\texttt{rakibul.hasan@asu.edu}\\
Arizona State University
}
\begin{document}
\maketitle

\begin{abstract}
Digital technologies have become ubiquitous in education, integrated to nearly all aspects of teaching, learning, research, and communication. They collect enormous amounts of data from their users, primarily students. When procured institutionally, data collection and use are guided by legally binding contracts. However, recent studies suggest that apps found on the open web are increasingly being used for teaching and evaluation without going through a formal audit, raising concerns about privacy, security, and regulatory compliance. How prevalent such shadow IT use in education is, and what privacy and compliance risks they create remain unexplored.

%We address this gap through a \textit{multi-source evidence} survey-based study: 
We address this gap through a sequential survey-based study: one survey with $375$ educators and another with $21$ system administrators at K-12 or HEIs (Higher Education Institutes). We discover widespread use of external apps by educators: $1,373$ apps were reported, $494$ among them were unique, primarily used for pedagogical utility $(n=213)$ and functional convenience $(n=155)$. Their adoption and integration are often facilitated, even \textit{encouraged}, by administrators; this result deviates from studies in other domains (e.g., industry) where admins actively prevent them. Many educators ($n=62$) continued using apps they do not trust, citing a lack of alternatives or dependency as barriers to discontinuation. We also found that fewer than one in three educators were aware of institutional policies on shadow IT use (K–12: $30.3\%$, HEI: $24.8\%$), and $22$ knowingly violated such policies. While $107$ received formal warnings for using shadow IT, only $33$ adjusted their behavior. We conclude by discussing the implications of our findings and future recommendations for various stakeholders to minimize the risks.
\end{abstract}

% keywords can be removed

\section{Introduction}

The ubiquity of digital technologies in education introduces new security and privacy risks. For instance, large data troves generated by tools for course management, teaching and grading, and communication have made this sector a prime target for cyber attacks~\cite{education_sector_risk, us-schools-data-breach, pearson-data-breach, nyc-data-breach1}.
Simultaneously, the lax privacy laws in the US for protecting student data---such as FERPA~\cite{ferpaLaw}, which is considered long outdated~\cite{russell2018transparencystudentdata, zeideLearnerPrivacyMOOCs2018, paris2022sins}---
incentivize direct selling or otherwise misusing students' data by technology vendors~\cite{student-data-sold, markup-farm-amassing, student-data-broker, zeideLearnerPrivacyMOOCs2018}. 

To minimize privacy and security risks, technology procurement at educational institutions follows a structured process that typically involves vendor evaluation, contract negotiation, and compliance review before procuring a new tool or service~\cite{kelso2024trust, chanensonUncoveringPrivacySecurityK12CHI23, shioji2025s}. Institutionally procured tools usually offer stronger security measures~\cite{CanvasAccountComparison2026} and privacy guarantees (e.g., not using customer data for product development~\cite{EnterprisePrivacyOpenAI1}) than their free or personal-use version. Crucially, institutional procurement creates legally binding contracts that serve as ``guardrails'' and enable institutions to hold vendors accountable for intentionally lax security measures or data misuse~\cite{kelso2024trust}.

However, recently, the use of software, hardware, or services without institutional procurement or authorization has begun to surface in educational contexts~\cite{levin2021state}. These \textit{shadow IT}~\cite{silic_shadow_2014, walterbusch_missing_2017}, i.e., technologies without formal audit or approval by the IT department, worsen the already precarious situation since they lack institutional oversight or contractual guardrails~\cite{van_acken_poster_2023, kelso2024trust}.

While the use of \textit{shadow IT} and associated risks (e.g., data loss, weakened security postures, and compliance violations~\cite{van2024department1}) has been explored in corporate settings (\S~\ref{sec:use}), research on technology adoption in educational contexts has largely centered on institutionally acquired tools. A couple of studies found evidence of educators using shadow IT~\cite{cohney_virtual_2021, chanensonUncoveringPrivacySecurityK12CHI23}; however, a comprehensive investigation of educators' role as everyday decision-makers who introduce \utech into teaching and evaluation is missing. We argue that educators, as the \textit{de facto} technology acquirers, make decisions that carry privacy and security implications for students and institutions alike. As data about students and minors are protected by specific laws in the US (\S~\ref{sec:regulations}), use of non-compliant tools can put institutions at risk of legal action and reputational damage. Mitigating this risk requires effective governance frameworks and institutional policies that are responsive to educators' needs, practices, threat perceptions, and the consequences of their decisions.

To contribute toward this goal, we investigated how and why educators in the US acquire and use \utech, drawing directly from their own perspectives; we complement this by examining the consequences of such use from the perspectives of system administrators.
Specifically, \textit{we conducted a survey study involving both educators and administrators to build an empirical foundation through complementary perspectives.}

% We focus on how  
First, we surveyed educators across K–12 schools and higher education institutes (HEI) to understand how they select and use tools for teaching-related activities (including applications and devices) outside institutional approval channels. We further examined educators' awareness of data privacy and regulatory compliance, and how personal factors (e.g., privacy literacy) and institutional culture impact their use of \utech. Building on these findings, we then surveyed system administrators, who work closely with educators, to understand how institutions perceive and respond to these practices. The following research questions guided these studies: 

\smallskip
\noindent\textbf{RQ1}:~How and why do educators in K–12 and higher education institutions adopt and use shadow IT?
% How (and why) are unsanctioned technologies adopted, and used by educators in K–12 and HEIs?

\noindent\textbf{RQ2}:~How do institutional and personal factors influence \utech use? 

\noindent\textbf{RQ3}:~What are the privacy and security implications of \utech use?

\smallskip
By exploring the above questions, we provide insights into the landscape, main drivers, and consequences of~\utech~use. We highlight the key findings below:

\smallskip
\noindent\textbf{Educators use apps and services not formally acquired by their institutions.}
In our survey, $375$ educators reported using $1,373$ apps ($494$ unique) (\S~\ref{edtech_use_prevalence}). 
Driving their initial adoption was \emph{pedagogy} ($n = 213$) and \emph{utility} ($n = 155$) (\S~\ref{edtech_selection_criteria}),
while continued use was justified predominantly by citing usability ($n = 294$). Additionally, $83\%$ of educators reported using personal devices for teaching, with $231$ storing student data on them and $127$ having personal cloud backups of those devices, raising significant privacy and legal compliance concerns.

\smallskip
\noindent \textbf{Security \& privacy aspects are rarely considered when adopting shadow IT, and its use may undermine the institutional security posture.}
Only $24$ educators explicitly mentioned \emph{security or privacy} as selection criteria. %Educators first-hand experienced privacy-invasive behaviors like excessive data collection and third-party data sharing, which led to discontinuation of apps for many ($n=167$)  (\S~\ref{observed_risks}), while some ($n=62$) continued their usage. %Some educators ($n = 62$), however, continued using apps after encountering similar behaviors, primarily due to a lack of alternatives. 
Our second study surveyed $21$ system administrators at educational institutions and identified $16$ serious concerns posed by the use of \utech (\S~\ref{sec:it_observed}), with \emph{data leakage} ($n = 13$) being the most cited risk. These risks are compounded when apps require students to register with institutional credentials, exposing institutional data to potential violations of federal regulations (\S~\ref{sec:privacy:compliance}).

\smallskip
\noindent\textbf{Institutional policy and its enforcement.} 
Only three administrators reported that their institution clearly prohibits the use of \utech, while 13 mentioned a case-by-case treatment. Fewer than a third of educators were aware of any institutional policy on \utech~use (K–12: $31.9\%$, HEI: $24.8\%$), and $22$ \textit{knowingly} disregarded them (\S~\ref{institute_policy_edtech_use}) as policy enforcement was rare.  
Seven admins even recommended \utech to educators, citing resource constraints and governance gaps as barriers to control (\S~\ref{institute_culture_edtech_use}). This is a striking deviation from prior research in corporate settings where admins actively prevent \utech use~\cite{silic_shadow_2014, kopper2017perceptions}.

%\begin{comment}
    
\noindent Our contributions are summarized as follows.

\begin{itemize} [itemsep=-1pt
, parsep=1pt]

\item We designed a survey-based study with educators and administrators to shed light on the use of \utech in education. To the best of our knowledge, this is the first study to understand the prevalence and implications of \utech use in educational settings.

\item We discovered 494 unique apps, suggesting widespread \utech~use in educational settings. %We also found that top reasons for use included pedagogical utility and functional convenience. 

\item We found that educators perceive apps as intrusive and doubt the developers' ability to secure user data; however, they continue to use \utech due to their deep dependencies.

\item We discuss how \utech degrades student safety in the short and long terms, and may lead to institutional non-compliance.

\end{itemize}

%\end{comment}
\section{Background and Related Work}

\subsection{Technology use in education}
\noindent
\textbf{Institutional adoption of technologies.}
Digital technologies have become ubiquitous at US educational institutions~\cite{kelso2024trust, chanensonUncoveringPrivacySecurityK12CHI23}. When institutionally acquired, they typically go through pre-deployment security and privacy audits, and their post-deployment behavior (e.g., what data they can collect) is constrained with legally binding contracts ~\cite{kelso2024trust}. While these tools have security and privacy implications~\cite{cohney_virtual_2021, chanensonUncoveringPrivacySecurityK12CHI23, balash2023educators}, this sanctioning process provides a layer of protection for students' data from being misused. 

\noindent
\textbf{Shadow IT in education.}
A handful of past studies reported educators' use of tools outside of institutional procurement process, particularly during and following the COVID-19 pandemic~\cite{levin2021state, cohney_virtual_2021, chanensonUncoveringPrivacySecurityK12CHI23}. In a recent short article, Acken \textit{et al.} listed \utech, along with associated risks (such as lack of access control and visibility to threats), commonly found in Dutch higher education institutes based on interviewing $11$ IT professionals~\cite{van_acken_poster_2023}. However, the impact of \utech use in the US context is largely unknown.

\subsection{Relevant federal and state regulations in the USA}\label{sec:regulations}

 Unfortunately, US lacks a federal privacy law. Some sector-specific laws dictate how technology accesses students' data, but their coverage depends on factors like the institution (public or private), student age, or \textit{type} of data.  For example, the \emph{{Children's Online Privacy Protection Act (COPPA)}.} protects personal information (e.g., name, address, biometrics) of children under 13~\cite{coppaa}. %Thus any technology serving at K-8 (kindergarten grade 8~\cite{ed_landscaoe}) schools most likely should comply with it. 
% Schools in the US, encompass kindergarten through 12th grade (referred to as K-12) and generally correlates to student ages 4--18 years. K-12 years are typically divided into three main parts, elementary school (K-5, with ages between 4 to 11), middle school (6-8, with ages 11 to 14), and high school (9-12, with ages 14 to 18) \cite{ed_landscaoe}. Thus any technology serving at K-8 schools most likely should comply with it. 
The \emph{{Family Educational Rights and Privacy Act (FERPA)}} covers Personally Identifiable Information (PII) and educational records for students of all stages~\cite{ferpaLaw}, %. The US Department of Education recommends grades, transcripts, class lists, student course schedules, health records (at the K-12 level), student financial information (at the post-secondary level), and student discipline files to be included under \textit{educational records}~\cite{educationRecord, ferpa-faq}, 
but what data is actually covered depends on the institutions~\cite{russell2018transparencystudentdata}. 
% Finally, the \emph{{Health Insurance Portability and Accountability Act (HIPPA)}.} applies to tools for health services provided by a covered entity (e.g., a hospital). Thus, it covers personal data of students receiving health services from hospitals or clinics operated under educational institutions~\cite{hipaa}.
These laws, particularly FERPA, have been criticized for being limited, vague, and containing loopholes that are regularly exploited~\cite{zeide2015student, paris2022sins, skowronski2022coppa}. %Moreover, FERPA and HIPAA typically do not apply to private entities like app developers unless they go into a contract with schools or health service providers~\cite{ferpa-faq, hipaa-definition}. Thus, due to weak regulations and the lack of contractual obligations, \utech are likely to pose privacy and safety risks to students. Moreover, educational institutes that receive federal funding can still be held liable in case of a security incident caused by \utech, and stands to loose funding and reputation.

% \textbf{State regulations.} Some states have developed additional laws to protect student data. Notable among them are the California Student Online Personal Information Protection Act (SOPIPA) enacted in 2016. Protecting data collected by education technology vendors about K-12 students, regardless of contracts vendors have~\cite{sopipa}. Illinois followed SOPIPA in creating their Student Online Personal Protect Act in 2019, that applies to technology vendors in contract with Illinois schools~\cite{illinois}. In 2017, Texas passed the Student Privacy Act which provides stricter protection to student data, and notably, prohibits targeted advertising for Texas students~\cite{texas}.
Thus, the weak and non-uniform regulatory landscape may allow technology vendors to misuse student data, and in states with stricter laws use of shadow IT may result in non-compliance and punitive actions.

\subsection{Research gap and motivation}
\label{sec:use}

Past research investigated \utech use in corporate settings (e.g.,~\cite{van2024department1, silic_shadow_2014, haag_acceptance_2019, kopper2017perceptions}); but a systematic study is lacking in the US education domain, which differ from corporate settings. The former is more exploration and innovation-friendly, and collaborative~\cite{kelso2024trust}, all of which might contribute to their adoption without guardrails. 
Acken~\etal~\cite{van_acken_poster_2023} provides a glimpse of privacy implications of \utech in educational institutions in the Netherlands. However, the Netherlands, operating under the strictest privacy law in the world~\cite{voigt2017eu}, differs significantly from the US both in legal environment and cultural norms~\cite{vitakWhenDataCollection-2023}. We hypothesize that these differences influence the adoption and use of \utech. Given the privacy and legal implications of their use, investigating their prevalence in the US education sector is critical.
\section{Methods}
%We investigate the landscape of~\utech, factors affecting their adoption, and associated privacy and compliance issues at US educational institutions. 
We adopt a sequential survey methodology~\cite{creswell2017designing}, comprising two surveys (the second informed by results from the first), to understand the prevalence (RQ1), practices (RQ2), and consequences (RQ3) of~\utech use in education contexts. %We developed two surveys to understand the views of shadow IT usage from our participant groups.

\subsection{Educator survey}
\noindent\textbf{\textit{Survey design.}} 
This survey was designed to gather both \textit{quantitative} and \textit{qualitative} insights from educators about the use of shadow IT. It comprised four sections: (1) app-specific behaviors and decision-making, (2) personal device use and data handling practices, (3) institutional policy and behavior changes due to apps' \sp, and (4) \sp behavior. Response options varied (e.g., multiple-choice, Likert-scale, and open-ended) depending on the question. The full questionnaires can be found at \url{https://doi.org/10.5281/zenodo.20385941}.

\textit{The first section} defined \textit{shadow IT} as \emph{any tool instructors acquired themselves and were not provided by their institution}. Participants had to indicate (by checking a box) that they read and understood the definition and had experience using them to proceed to the next question. Next, participants listed three to five such apps that they have used for teaching related activities. Follow up questions for each app asked about selection motivation, discovery sources, integration with institutional tools, requirements of student registration with institutional credentials, current use status, and regulatory compliance awareness. 
Educators were also asked about privacy issues they observed during usage, trust in the app and developers, and presence and utilization of any data deletion options when apps were discontinued. \textit{The second section} asked about using personal devices for work, e.g., downloading student data to personal laptops or phones, and whether those devices had automatic cloud backups. \textit{The third section} explored participants’ awareness of institutional policies, how they were communicated, and if they had caused changes in their app use. \textit{The fourth section} included psychometric scales to measure participants' attitudes toward protecting other people's privacy (the VOPP scale~\cite{hasan_psychometric_2023}) and security habits (the SeBIS scale~\cite{Egelman2015ScalingSeBIS}). The survey ended with demographic questions about age, gender, and teaching level (K-12 vs HEI).

\smallskip
\noindent\textbf{\textit{Recruitment.}}
The survey was implemented on Qualtrics~\cite{Qualtrics} 
and advertised on Prolific~\cite{prolificReseach}. 
We used Prolific's prescreening option~\cite{HowUseProlifics1} to recruit crowdworkers who are current or former educators at K-12 or HEIs in the US. The survey consent form explicitly asked about prior experience using \utech (since we aimed to explore experience and associated privacy issues) and allowed only those who answered affirmatively to participate. We piloted the survey ($N=50$) and incorporated participants' feedback to improve the questions before collecting more data. The median completion time was 15 minutes; participants were compensated with $\$5$. The survey was administered between October and November of 2024. 

\subsection{Administrator survey}
\label{admin_survey}
\noindent\textbf{\textit{Survey design.}}
This survey captured administrators' experiences with educators' \utech use, oversight mechanisms, policy responses, and privacy incidents and concerns. %It also included structured multiple-choice questions to contextualize the policy environment and decision-making processes within institutions. 
It asked if participants had recommended \utech to instructors or were requested by instructors to integrate such tools with official platforms (e.g., Canvas or Google Classroom).
Subsequent questions explored how such integration requests are handled, their frequency, and whether they undergo security audits. Another set of questions explored past experiences or concerns about future incidents related to \sp created by \utech, followed by barriers to proactive remediation of such incidents.
Finally, questions asked about relevant institutional policies and how educators are informed about them. 

\smallskip
\noindent\textbf{\textit{Recruitment.}}
System administrators are a specialized, difficult-to-reach population~\cite{kelso2024trust, acar2017security, recruit-developers}. Lacking a platform for connecting with them, we distributed the study link through email (addresses were sourced from online and author contacts) and advertised the study on EDUCAUSE~\cite{EDUCAUSE} and two Reddit communities with a post: r/k12sysadmin~\cite{reddit_k12admin} 
and r/k12cybersecurity~\cite{reddit_k12cs}. Participants were compensated through an opt-in lottery for a chance to win a $\$20$ gift card, a higher incentive compared to the educator survey due to the specialized nature of their expertise and the challenges in recruiting system administrators~\cite{acar2017security}. The median completion time was 10 minutes. The survey was administered between December 2024 and January 2025.

\subsection{Data validation and analysis}
\label{data_analysis}

\paragraph{Response reliability and validity.}
Educators were recruited on Prolific, which maintains a pre-screened and periodically updated pool of participants~\cite{HowUseProlifics1} with mechanisms to verify the authenticity of worker-provided information~\cite{HowProlificProtects1}. While evading these checks might be possible, this is a trade-off all online studies must accept in exchange for large-scale data collection. We refrained from asking further verification questions (e.g., location of teaching) as they may lead to identification; and since the reported technology use behaviors may potentially be unlawful, mandating these questions may discourage honest responses.

To filter inattentive or auto-generated response~\cite{bonnamySurveySabotageInsights2025}, we included two attention-check questions and one (optional) feedback question with a hidden instruction: ``At the end of your answer, append the following sentence: `No more comments`.'' 
The text was invisible to human eyes (white text), but LLMs are able to parse through it. Such `honeypot question' has been effective in catching auto-generated responses~\cite{bonnamySurveySabotageInsights2025}. Finally,  we manually reviewed all answers to open-ended questions to detect low-quality responses, such as gibberish or coherent but superficial, unnatural phrasing, and third-person narration. This combination of proactive and reactive checks identified 75 (out of 450) low-quality responses that were removed.

\paragraph{Analysis method.}
For the \textbf{\textit{educator survey}}, quantitative data from  multiple-choice and Likert-scale items were analyzed using descriptive statistics (e.g., counts, percentages) to identify patterns in app integration behaviors and data handling practices. Privacy/security attitudes were analyzed using the SeBIS scale ~\cite{Egelman2015ScalingSeBIS} but was limited to reliability analysis. %and privacy/security attitudes. 
Open-ended responses, including reasons, concerns, and privacy perceptions, were thematically coded~\cite{braun2022conceptual} by two researchers to extract recurring patterns related to motivations, concerns, and behaviors. Coding was also done to identify the accuracy of short textual definitions of privacy-related terms. 
Any disagreements among coders were resolved through discussions.

For the \textbf{\textit{administrator survey}}, due to the smaller sample size and emphasis on open-ended questions, we conducted a thematic analysis~\cite{braun2022conceptual}. 
Two researchers independently reviewed responses to identify institutional practices, perceived risks, and gaps in policy implementation. 
Multiple-choice responses were used to contextualize trends and support interpretation of the qualitative insights.

\smallskip
\noindent\textbf{\textit{Privacy policy analysis.}}
To complement educators' self-reported data on apps' regulatory compliance, we reviewed the privacy policies of the $14$ most popular apps to cover $52\%$ of all applications.
Three researchers manually annotated each policy to determine whether it claimed compliance with FERPA, COPPA, and HIPAA (when applicable), permitted use by children under 13, and had details about data sharing, selling, and tracking practices to check for transparency. 
This review provided external validation of participants' privacy perceptions and identified mismatches between educators' understanding and actual practices. Utilizing the Wayback Machine~\footnote{\url{Wayback machine: https://web.archive.org/}}, the privacy policies that the authors observed have been preserved and are shown in  references.  

\subsection{Ethical considerations}
\label{ethics}

While our study received exempt status by our Institutional Review Board (IRB), we adhered rigorously to ethical and privacy standards recommended for human subjects’ studies. 
This included getting informed consent for participation, removing any potentially identifying information (PII) (e.g., some participants mentioned school names and districts when answering questions), and ensuring that all collected data was stored securely and only used for research purposes. Since information present in qualitative responses may leak identity and the participants discussed technology use that could violate institutional policies, we will not make the dataset publicly available. However, upon request and after reaching an agreement over confidentiality, we will provide data to other researchers.

\subsection{Study limitations}
\textit{First}, we focus on the US; findings may not generalize to other countries with different regulatory environments. We recruited educators from an online platform, requiring the usage of shadow IT, thus lacking the perspectives of those who never used those technologies; limiting the generalizability of the findings. Likewise, our methods to reach admins might have preferred more security-conscious participants than the general population. This cohort, however, is also more likely to have handled \sp incidents arising at their institutions, enriching our data. On the other hand, many administrators in our sample allowed, even encouraged \utech use; such findings indicate that \utech may enjoy more leniency in general.
%; the participants might be more tech-savvy than the general population of educators, potentially impacting the generalizability of the findings. Likewise, our methods to reach administrators might have preferred more security-conscious participants than the general administrator population. This cohort, however, is also more likely to have handled \sp incidents arising at their institutions, enriching our data. On the other hand, many administrators in our sample allowed, even encouraged \utech use; such findings indicate that \utech may enjoy more leniency in general.

\textit{Second,} interpreting open-ended responses may introduce subjectivity, a common challenge in qualitative research. We applied collaborative theme refinement and double-coding to reduce subjectivity and improve consistency. 

\textit{Third,} some questions required participants to remember prior decisions or experiences. Answers may suffer from a recall bias, a common issue in studies like ours. We anticipate its impact to be minimal as our survey focused on current and regular behaviors.
% , which could create recall bias and is a commonly observed limitation in data collection methods of surveys and interviews.
% To improve recall accuracy, we phrased questions clearly and encouraged concrete examples
% \SR{Is there any supporting evidence that it actually improves? If not, I would remove this limitation ...}\easton{i dont know.. ananta?}.

\textit{Fourth,} as stated we did not include participants who never used shadow IT. This was a deliberate design choice as we aimed to understand the experience of the users. Additionally, we asked participants to list at least three apps they used (to capture diversity), which could artificially inflate the numbers; however many participants voluntarily listed more than three apps, and the total number of unique apps was high. Thus, the biasing impact of this question seems minimal.

\textit{Lastly,} as recruiting participants from such a specialized field is challenging (\S\ref{admin_survey}), we employed multiple outreach strategies to reach system admins, as well as offering higher compensation through an opt-in lottery. 
Despite these efforts, our final sample size remained relatively small and the distribution between K-12 and HEI settings was uneven. However, we ensured data quality and relevance by screening responses for completeness and authenticity, allowing us to capture meaningful insights from these participants.

Despite these limitations, this study provides a strong empirical foundation for understanding \utech~use. Findings reported here will guide future research exploring mitigation for privacy and security issues from \utech. %exploring  these dynamics in broader and more diverse populations to validate and extend our results.

% \section{Results}
% Based on our analysis of two online-surveys with educators and administrators from K-12 and HEI settings, we now present our study's findings and start with describing our participants and their demographics.

\subsection{Study participants}
\noindent\textbf{Educators.} A total of $450$ participants took our survey on Prolific.
Of these, $75$ responses were excluded after a manual review, finding them either auto-generated, gibberish, or duplicates. 
The remaining ($n=375$) responses were used for analysis: $251$ from K-12 educators and $124$ from HEI educators. Throughout the paper, we refer to educator participants using the prefix `P' followed by a unique ID (e.g., P12).
Our participants represented diverse demographics: $260$ identified as female, $110$ as male, and $5$ as non-binary. 
The age distribution was as follows: $83$ participants were under age  $30$ (18–29), $225$ were 30–50, and $67$ were over $50$.
Regarding education, $18$ had a high school diploma, $70$ had a Bachelor’s degree, $65$ had a Master’s degree, $12$ had a doctorate, and the rest preferred not to answer.
%Discipline-wise, $20$ participants taught STEM subjects, $16$ taught non-STEM subjects, $12$ taught in the Education field, four were from Business, three from Health-related fields, and five participants fell into Other categories.

\medskip
\noindent\textbf{System Administrators.} To correlate our findings with educators, we surveyed $21$ administrations, denoted with an `A', to discuss the experiences they shared with~\utech~ and its consequences. Administrators were not necessarily IT personnel but also individuals in positions of power over the institution such as vice principles.
Among those, two were from K-12 institutions, while the rest were from HEIs. Of those participants, twelve were actively in roles related to information security, and the rest were in administrative roles at their institutions, such as business manager, principal, or vice chancellor. 

\newcolumntype{Y}{>{\centering\arraybackslash}X}
\section{Shadow IT Landscape in Education (RQ1)}
\label{rq1_results}

This section describes results from the educator survey. Based on both quantitative and qualitative responses, it portrays the landscape of \utech use in K-12 and HEI settings. Since many questions in the survey were related to or follow-ups of others, we present results from multiple questions together to better contextualize and create an easy-to-follow narrative.
%Within our sample, \utech is dominated by apps and services, but use of personal devices were also commonplace. In the following sections, we detail how educators discover and select apps and services, identify and categorize most popular them, and end with the different ways they use personal devices for professional purposes.

% \rakib{Do not use CamelCase for subsections}

\subsection{Prevalence and categories}
\label{edtech_use_prevalence}

Below we report educators' use of apps and devices outside of institutional oversight.
% To map the landscape of~\utech, educators were first asked to \emph{list three to five apps they have used that had not been formally approved by their institutions} (educator survey, Q.\ref{application list}). Another question covered how they use personal devices to conduct \textit{institutional} activities (educator survey, Q.\ref{device_use}). Here, we report results about apps followed by device use.

\smallskip

\noindent
\textbf{Applications and services.}
In total, participants listed $1,373$ apps and services ($494$ were unique), highlighting their ubiquity in education. 
A majority of the participants ($n=213$) listed three apps, $73$ listed four, and $88$ listed five.
Among the $494$ unique apps, $285$ were listed only by K-12 educators (top three: Google Classroom, Kahoot!, and Canva), $121$ were listed only by HEI educators (top three: ChatGPT, Kahoot!, and Canvas), and the remaining $88$ were listed by both groups.\footnote{Note that popular tools like Canvas may be institutionally procured at many institutions, but not all; in those institutions it would be considered as a shadow IT.} These results highlight both shared trends and context-specific needs across sectors. Table~\ref{tab:top_apps} presents the most commonly cited apps, which were categorized on two fronts: 1) \emph{pedagogy-focused} (e.g., tools mainly used for teaching, assessment, or content creation) and \emph{utility-focused} (e.g., tools for communication, content delivery, or storage). %\SR{Add a rough definition of \textit{pedagogy} and \textit{utility} with some example}.

% Suggestion by Reviewer: there is wide variety in the unsanctioned apps that educators use.
%\SummaryBox{ The large number of apps educators listed, and the fact that many listed more than three apps, indicate that \utech use has become widespread.}

Alarmingly, some of the apps still in use either are no longer supported by their developers, or the companies behind them declared bankruptcy. Edmodo~\cite{edmodo_app}, for example, was listed by three educators; only one has discontinued its use, despite the company shutting down in 2022 due to violating COPPA and being sued by the Federal Trade Commission (FTC). Blackboard~\cite{blackboard-phase-out} has phased out support for self-hosted deployments, potentially affecting individuals relying on it. Additionally, ChatGPT's terms of service prohibit use by minors under 13, making its use in the K–12 sector concerning.

%\SummaryBox{
%Some educators continue to use \utech even when it is legally challenged by government regulators or abandoned by the vendor.}

\begin{table}[h]
    %\footnotesize
    % \renewcommand{\arraystretch}{1.2} % Default value: 1
    \small
    \caption{Most commonly listed shadow IT applications with their primary category and frequency of educators (both K-12 and HEI) listing the app use.}
    \label{tab:top_apps}
    \centering
    \begin{tabular}{p{4.7cm} p{3cm} p{1cm} p{1cm}}
        \midrule\midrule
        \textbf{Application} & \textbf{Category}& \textbf{\# K-12} & \textbf{\# HEI} \\
        \midrule \noalign{\vspace{2pt}}
        Kahoot! & Pedagogy & 42 & 18 \\ \noalign{\vspace{2pt}}  \hdashline \noalign{\vspace{2pt}}
        ChatGPT & Utility & 31 & 24 \\ 
        \noalign{\vspace{2pt}}  \hdashline \noalign{\vspace{2pt}}
        Google Classroom & Utility & 43 & 12 \\ 
        \noalign{\vspace{2pt}}  \hdashline \noalign{\vspace{2pt}}
        Canva & Pedagogy & 33 & 14 \\  
        \noalign{\vspace{2pt}}  \hdashline \noalign{\vspace{2pt}}
        ClassDojo & Utility & 31 & 5\\ 
        \noalign{\vspace{2pt}}  \hdashline \noalign{\vspace{2pt}}
        Quizlet & Pedagogy & 23 & 11 \\ 
        \noalign{\vspace{2pt}}  \hdashline \noalign{\vspace{2pt}}
        Canvas & Utility & 18 & 15 \\  
        \noalign{\vspace{2pt}}  \hdashline \noalign{\vspace{2pt}}
        Teachers Pay Teachers & Utility & 22 & 3\\ 
        \noalign{\vspace{2pt}}  \hdashline \noalign{\vspace{2pt}}
        Youtube & Utility &15 & 9\\ 
        \noalign{\vspace{2pt}}  \hdashline \noalign{\vspace{2pt}}
        Zoom & Utility & 11 & 9\\ 
        \noalign{\vspace{2pt}}  \hdashline \noalign{\vspace{2pt}}
        Grammarly & Utility & 9 & 11\\ 
        % \noalign{\vspace{2pt}}  \hdashline \noalign{\vspace{2pt}}
       \bottomrule
    \end{tabular}
\end{table}

\smallskip
\noindent
\textbf{Personal device use.}
\label{personal_usage}
A large majority of educators ($83\%$) reported using personal devices (e.g., laptop, tablet, smartphone) for teaching-related tasks daily ($n=93$) or several times per week ($n=112$). Educators ($n=231$) also reported downloading institutional documents---such as grade books or student records---onto personal devices. Among them, $127$ had automatic cloud backups enabled through personal accounts; this could introduce compliance risks since FERPA requires student data storage to be compliant~\cite{ferpaLaw}.

\subsection{App discovery mechanisms}
\label{edtech_discovery_mechanism}

% To discover unofficial sources for shadow IT, we asked educators where they had learned about the apps they had listed (Q.\ref{app_learn}). Table~\ref{tab:discovery_mechanisms} shows the results. 

Shadow IT was discovered mainly via personal and professional connections (Table~\ref{tab:discovery_mechanisms}); %This ecosystem of trust---in colleagues, friends, institutions, and online platforms---highlights the influence of informal channels in app discovery and adoption, bypassing formal approval processes.
this echos prior work on how individuals turn to familiar, non-expert sources for technology and security guidance~\cite{redmiles2016learned, redmiles2016think}, reinforcing the notion that trust-based, informal networks are a key driver in the adoption of~\utech.

\begin{table}[h]
    % \footnotesize
    % \renewcommand{\arraystretch}{1.2} % Default value: 1
    \small
    \caption{Prevalence of shadow IT discovery mechanisms reported by educator participants}
    \label{tab:discovery_mechanisms}
    \centering
    \begin{tabular}{p{6cm} p{1cm}}
        \midrule \midrule
        \textbf{Discovery Mechanism} & \textbf{Count} \\
        \midrule
        Professional Peers & 290 \\ \noalign{\vspace{2pt}}  \hdashline \noalign{\vspace{2pt}}
        Admins or Supervisors & 141 \\
        \noalign{\vspace{2pt}}  \hdashline \noalign{\vspace{2pt}}
        Online Search or Browsing & 100 \\
        \noalign{\vspace{2pt}}  \hdashline \noalign{\vspace{2pt}}
        Advertisements or Sponsored Content & 100 \\  
        \noalign{\vspace{2pt}}  \hdashline \noalign{\vspace{2pt}}
        Friends or Family & 44 \\  \noalign{\vspace{2pt}}  \hdashline \noalign{\vspace{2pt}}
        Social Media or Influencers & 31 \\ \noalign{\vspace{2pt}}  \hdashline \noalign{\vspace{2pt}}
        Workshops, Conferences, or Webinars & 29 \\   \noalign{\vspace{2pt}}  \hdashline \noalign{\vspace{2pt}}   
        Previous Experience / Already Familiar & 4 \\ 
        % \noalign{\vspace{2pt}}  \hdashline \noalign{\vspace{2pt}}
        % Miscellaneous or App Features Related & 634 \\
        \bottomrule
    \end{tabular}
\end{table}

%\SummaryBox{
%Non-security experts, primarily professional peers or supervisors and online platforms, play the biggest role in \utech discovery.}

\subsection{Rationale for selection and continued use}\label{edtech_selection_criteria} 
Educators' provided theirs reasons for using the apps they listed; they were manually coded, yielding $17$ unique causes grouped into six categories summarized below and listed in Table~\ref{tab:app_factors_category}.

\smallskip

\noindent
\textbf{Rationale for selection.}
\emph{Pedagogy} was the most common reason ($n = 213$) for selecting apps, including those used for instruction, assessment, and content delivery. 
\emph{Utility} ($n = 155$) followed, dominated by apps used for grading, lesson planning, communication, and learning management. 
Many apps spanned multiple subcategories; e.g., Magic School AI~\cite{magicschool} provides assessment and AI-driven personalization features.
Several apps under \emph{pedagogy} were selected for \emph{gamified learning} ($n = 30$). Twenty apps under \emph{utility} were selected for their AI features, offering utilities for automated grading, content editing and creation, and personalized learning plan development.
% This indicates that while AI offers promising advancements in education, it also presents risks like over-reliance, data privacy concerns, and potential biases~\cite{AI-education-risks}, requiring careful consideration during selection.
During selection, educators prioritized \emph{accessibility} ($n = 187$) and \emph{cost} ($n = 164$), often preferring free, easy-to-access apps.
Many participants ($n=117$) opted for shadow IT applications even when institutional alternatives were available. Most cited causes behind this were: \textit{(i)} familiarity or better user experience with the shadow IT, or \textit{(ii)} institutional tools being outdated or poorly maintained. Finally, only 24 out of 375 participants explicitly mentioned  security or privacy as a selection criterion. %``security,'' ``privacy,'' ``secure,'' ``safe,'' or ``safety'' when expressing their selection criteria, indicating that concerns about data protection were not prominent during the initial app adoption phase for educators. %indicating that concerns about data protection were largely absent from most educators' EdTech-related decision-making processes.

%\SummaryBox
%{Familiarity often influenced \utech selection over institutionally-approved alternatives and far eclipsed security and privacy considerations.}

% Among the $177$ participants who acknowledged having an institutional alternative, 84 chose to stick with their preferred tools due to previous careers, long-term use, and unwillingness to adapt to a new tool.

% Some applications did not fall into our created categories and
% were put under miscellaneous, this includes Ad Blocker Extension,
% timers, and Voice-to-text Extension.
% \smallskip

\noindent
\textbf{Rationale for continued use.}
% After initial adoption, educators often made a deliberate choice to continue using~\utech. 
%We asked educators to state reasons for continuing to use an app after adoption (Q.\ref{app why} and Q.\ref{still_in_use}). Their responses were manually coded to identify patterns. 
\emph{Ease of use} was the most commonly cited reason ($n = 294$), followed by \emph{improved student engagement} ($n = 108$), which was noted by $65\%$ of K-12 educators.
In HEI contexts, the need for specialized features drove continued use ($n=61$). %Educators cited needs ranging from \qoute{``basic image editing''} (P300), \qoute{``reference apps for specific films and developer combinations''} (P151), to \qoute{``help with research papers''} (P52), or \qoute{``coding observational data''} in lab environments (P37).
%

% there is wide variety in the unsanctioned apps that educators use.

\SummaryBox{\textbf{RQ1 Summary:} Practical needs drive shadow IT adoption where security/privacy considerations are often not prioritized---raising critical questions about risk awareness and compliance.}

\begin{table}[t]
\small
\centering
% \footnotesize
\newcolumntype{Y}{>{\centering\arraybackslash}X}
\caption{Categories of apps, based on \textit{cited} rationales.}
\label{tab:app_factors_category}

\begin{tabular}{p{5cm} p{6cm} p{1cm}}
\midrule \midrule
\textbf{Category} & \textbf{Subcategories} & \textbf{\# Apps} \\
\hline
\noalign{\vspace{2pt}}
\multirow{10}{*}{Pedagogy}
 & Teaching Aides & 52 \\
 & General Ed & 31 \\
  & Gamified Learning  & 30 \\
  & Math & 22 \\
  & Reading & 21 \\
   & English & 14 \\
   & Engineering & 13 \\
 & Assessments & 10 \\
 & Science & 10 \\
 & Special Education & 10 \\
 % & \textit{Total} & 213\\

\noalign{\vspace{2pt}}
\hdashline
\noalign{\vspace{2pt}}
\multirow{6}{*}{Utility}
 & Learning Support & 74 \\
 & Learning Management & 43 \\
  % & Research Support & 24 \\
  & AI Tools and Utilities & 20 \\
 & Specialized Items & 9 \\
 & Editors & 9 \\
\noalign{\vspace{2pt}}
% & \textit{Total} & 155 \\
\hdashline
\noalign{\vspace{2pt}}
\multirow{1}{*}{\centering Accessibility}
 & -- & 187 \\
\noalign{\vspace{2pt}} 
\hdashline
\noalign{\vspace{2pt}}
 \multirow{1}{*}{\centering Cost}
 & -- & 164 \\
 \noalign{\vspace{2pt}}
 \hdashline
 \noalign{\vspace{2pt}}
\multirow{2}{*}{\centering Familiarity}

 & Lack institutional tools & 84 \\
 & Long-term use & 33\\
 
\noalign{\vspace{2pt}}
\hdashline
\noalign{\vspace{2pt}}
\multirow{1}{*}{\centering Sec. \& Privacy}
 & -- & 19 \\
\bottomrule
\end{tabular}
\end{table}

\section{Impact of institutional and personal factors on \utech use (RQ2)}
\label{RQ3}
This section reports how institutional culture, including data privacy policy and enforcement, and educators' privacy attitudes, impact \utech use.

\subsection{Institutional culture}
\label{institute_culture_edtech_use}

\noindent\textbf{Privacy policy, their enforcement and awareness.}
\label{institute_policy_edtech_use}
Educational institutions maintain privacy policies that dictate the use of technology~\cite{kelso2024trust, chanensonUncoveringPrivacySecurityK12CHI23}. We asked administrators if their institutional policy clearly guide \utech use. Three administrators indicated that \utech was strictly prohibited, $13$ mentioned they decided on a case-by-case basis, and the rest were unsure. All but two administrators stated that these policies are being updated regularly, and all confirmed that policies were disseminated online (three mentioned they additionally sent emails to faculty). However, knowledge of policies among the educators was not widespread: only $31.9\%$ ($n = 80$) of K–12 and $24.8\%$ ($n = 30$) of HEI educators were aware of such a policy. Even among those who were aware, $22$ admitted to using shadow IT in violation of policy. As P10 noted, \qoute{``[My institution doesn't] provide a workable path to utilize new and emerging technologies, so [I] have to go against it.'' }
Similarly, P68 stated, \qoute{``We are not supposed to use any application that is [not] already approved, but many, like myself, ignore the rule.''} This practice was not secret: many administrators knew about instructors bringing external tools into the classroom. As A19 described, \qoute{``Almost at an individual level, everyone has their `solution' to teaching/learning `better' and brings it into the classroom with disregard.''} Thus, policy enforcement was rare, as echoed in administrators' frustrations---\qoute{``There is a policy, but it needs enforcement'' }(A8). A21 shared that policies were \qoute{``regularly ignored and rarely enforced.''} 

Many institutions, however, attempt to detect policy violations and raise awareness: 107 educators reported receiving institutional warnings about risks associated with shadow IT. Of those, $33$ stated that they modified their behavior by discontinuing the use of specific apps, switching to more secure alternatives, enabling two-factor authentication, or limiting what data they shared with third-party platforms.

%\SummaryBox{Educators are often unaware of \utech-related institutional policies, which are often minimally enforced or maintained. Only a small portion (33/107) of educators report modifying their behavior after receiving institutional warnings on using \utech.}

\noindent\textbf{Resource constraints and governance gaps.}
%When asked about proactive strategies to reduce risks, most administrators referenced existing guidelines or resources available to educators %(AQ.\ref{admin_proactive} to AQ.\ref{admin-barriers}).
Administrators cited several barriers to implementing technology use guidelines: cost, under staffing, and limited IT capabilities. 
% Three admins explicitly identified resource constraints as the main obstacle to proactive oversight.
% 
Some administrators framed the issue not only as a matter of compliance or risk but also as a structural governance failure. A14 noted: \qoute{``I don’t think this is a security and privacy issue---it’s a governance issue. Institutional leaders need to give clear guidance on when non-institutionally licensed tools can be used by instructors, who can make the decision about data used in those tools, and who can process exceptions. This isn’t an IT issue so much as it is a governance issue.''}

%\SummaryBox
%{Lack of resources and unclear guidance hinder administrators' ability to give clear answers on appropriate unsanctioned technology usage and the proper authority to halt such uses.}
%Permissive culture and approval from administrators are among the major drivers of \utech adoption.

\noindent\textbf{Administrators' informal endorsement.}
As mentioned above, %most institutions consider \utech use on a case-by-case basis, and generally, they are permissive. 
all administrators indicated that they were generally aware of instructors' use of shadow IT. Moreover, seven of them acknowledged recommending the use of such apps (AQ.$1$\footnote{Refers to the question number in the full questionnaire at \url{https://zenodo.org/records/20385941}}), often to \qoute{``meet niche demands that our institutional platforms cannot manage, or where instructors are looking for free alternatives for their use'' }(A4). 
Requests for integrating these apps into the institutional ecosystem were also widespread. %(AQ. \ref{admin_int}). 
All but three administrators had received such requests from instructors; e.g., A2 noted, \qoute{``We have instructors wishing to integrate many tools such as polling, scheduling, [and] citation software for use in classes.'}' Two participants said integration requests were never accepted, while others said they were accepted ``sometimes'' or ``most of the time.'' Twelve administrators reported that such requests required IT audits, while others were unsure whether audits were conducted (AQ.$6$).

%\SummaryBox
%{More often than not, \utech use is informally endorsed (and sometimes recommended or facilitated) by the IT administrators.}

%\SummaryBox
%{Low awareness and enforcement of shadow IT usage creates an environment where educators utilize tools that go against policy.}

\subsection{Educators' S\&P behaviors and attitudes}\label{sec:personalfactors} 
%\ananta{Easton, can we have exact p-values for everything that we are reporting here instead of p>0.05 or <0.05 ? I think it will provide stronger evidence for significance.}

%\easton{This section was a bit of a weak point from the reviews, we dont really have a great answer to why did we only use half of a scale, same with privacy literacy.}\ananta{Easton, could you find papers who have leveraged only a subset of the scale?} \easton{issue: the sebis scale has sub scales in it that have been proven; however we used it in a really weird way we did a sub scale of a sub scale???}
\label{educator_sp_attitudes}
We report educators' security and privacy behaviors and attitudes and their correlations with Shadow IT use.

\smallskip
\noindent
\textbf{General security behaviors.}
We used four items from the SeBIS scale~\cite{Egelman2015ScalingSeBIS} to measure general security practices among educators (Q.$13a$ to Q.$13d$). We did not use the full scale as many items are not relevant (e.g., locking screen) to our study, and including them would increase the study duration, risking burnout. We computed the final score by averaging individual item response. $96.8\%$ of participants scored $3.5$ or higher on a 5-point scale, and %, which is higher compared to the general public in the original study, as Table~\ref{tab:sp} shows in the Appendix. 
the difference between HEI and K–12 was minimal: $90.3\%$ of HEI educators and $100\%$ of K-12 educators scoring $3.5$ or higher. %, while $100\%$ of K–12 participants did.
Next, we tested if the SeBIS score correlated with privacy considerations during \utech selection, by performing t-tests. %\ananta{Can we add which test we performed?}. 
Specifically, we asked educators if they considered privacy or security during app selection; only $72$ educators responded affirmatively for all apps they used, and $176$ did so for at least one app. This group scored slightly higher on the SeBIS scale ($M = 4.23, SD=0.79$) than those who never considered privacy during app selection ($M= 4.03, SD=0.79$); but the difference was not significant ($p=0.052$). Thus, following general security advice may not directly correlate to shadow IT use.

\smallskip

\noindent\textbf{Attitude toward others' privacy.}
Since \utech uses primarily risks the privacy of other people (i.e., students), we used three items from the VOPP scale~\cite{hasan_psychometric_2023} to measure how much educators value other's privacy (Q.$13e$ to Q.$13g$). Since this is a single factor scale but we did not use all items, we computed the reliability score to see if the items can be used reliably and found a high Cronbach’s $\alpha=0.91$.%, \diff{indicating that the items coherently capture common underlying construct and can be combined into a composite score for analysis~\cite{hasan_psychometric_2023}}. %\ananta{A pets reviewer asked explain $\alpha$: ``We assessed the internal consistency of the selected items using Cronbach’s $\alpha$ and observed high reliability ($\alpha$ = 0.91), indicating that the items coherently capture a common underlying construct and can be combined into a composite score for analysis.''}
For these items, the sample mean was $4.57$ ($SD= 0.61$), suggesting educators' strong concern for others' privacy. Participants who considered privacy while selecting apps scored significantly higher  ($M=4.71, SD=0.49$ versus $M= 4.54, SD=0.62$, $p=0.0126$) than the rest, indicating that valuing others' privacy may lead to a more thoughtful selection process.
%Future study could investigate if high VOPP scores correlate with the avoidance of \utech and design interventions for raising awareness among instructors about students' right to privacy.

\SummaryBox{\textbf{RQ2 Summary:} A lack of awareness about policy and its lax enforcement contribute to shadow IT use. While general security behaviors did not correlate with shadow IT use, educators’ valuation of other people’s privacy was linked to considering privacy and security aspects during shadow IT selection.}

\section{Security and Privacy (S\&P) Implications of \utech Use (RQ3)}
\label{SP-implications}

% \SR{We need an opening sentence to connect this part with RQ1.}
This section reports on the security, privacy, and regulatory compliance aspects of \utech. 

\subsection{Educators'  S\&P experience}
\label{observed_risks}

\smallskip
\noindent
\textbf{Privacy invasive data collection and sharing.} 
We asked educators if they observed app behaviors that they considered invasive (\S~\ref{edtech_use_prevalence}, Q.$4m$ to Q.$4o$). $47$ participants indicated that they believed at least one app they used collected data they considered sensitive. Multiple participants indicated similar observations in apps; e.g., four participants mentioned about ChatGPT, three mentioned Blackboard, and YouTube and Canvas had two complaints each. Additionally, 62 participants suspected that the apps they used sold or shared user data with third parties: among them, five participants mentioned ChatGPT, four mentioned YouTube; each of Canvas, Class Dojo, and Quizlet was mentioned by two participants. %In with this, many still continued to use their applications.

\smallskip
\noindent
\textbf{Distrust in the technical competence of App developers.} 
Many educators expressed distrust in the technical ability of apps to safeguard user data (Q.$4r$). Thirty-eight participants explicitly stated that they distrusted at least one of their apps; one participant distrusted all of them. $31$ participants believed that at least one app's developers were not competent to protect user privacy, with one participant holding this belief against developers of all apps they listed.

Despite these concerns, all 62 participants who distrusted an app continued using that, citing reasons such as the apps being deeply embedded in their workflow or not having suitable alternatives. Ten of these participants not only continued using the app, but also had their students register with those apps (Q. $4e$); three of them also stated that those applications were assumed to be not FERPA compliant (Q.$4p$). These findings highlight a tension between perceived risk and pedagogical or practical utility that may outweigh privacy concerns in day-to-day decision-making.

%\SummaryBox{Educators overwhelmingly continue to use \utech despite their own concerns about the collection and misuse of students' data.}

\smallskip
\noindent
\textbf{App discontinuation.} We asked educators if they ever uninstalled an app because of privacy concerns (Q.$12g$ to Q.$12i$). 167 educators reported that they did for at least one app. Two authors categorized the open-ended responses about invasive behaviors leading to app discontinuation, as listed in Table~\ref{tab:Reasons}.

\smallskip
\noindent
\textbf{Data deletion after discontinuing apps.}
$94$ participants stated that at least one app they listed was not in use anymore (Q.$4f$). We asked them if they deleted data the apps had collected (Q.$4g$). Only $13$ participants reported that the app offered a data deletion feature and $9$ used it; the rest were unsure whether such a feature existed. It is noteworthy that the remaining $4$ participants who did not delete data also reported that they mandated students in their class to register for that application. 

\begin{table}[h]
\small
    \renewcommand{\arraystretch}{1} % Default value: 1
    \caption{Educators' security-related reasons for uninstalling apps.}
    \label{tab:Reasons}
    \centering
    \begin{tabular}{p{8cm} p{1cm}}
        \midrule \midrule
        \textbf{Reason} & \textbf{Count} \\
        \midrule
        The app asked for unusual permissions &  52 \\ 
        \noalign{\vspace{2pt}} \hdashline \noalign{\vspace{2pt}}
        The app collected more data than stated & 25 \\
        \noalign{\vspace{2pt}} \hdashline \noalign{\vspace{2pt}}
        Shared data with third parties & 17 \\
        \noalign{\vspace{2pt}} \hdashline \noalign{\vspace{2pt}}
        Tracked user & 15 \\  
        \noalign{\vspace{2pt}} \hdashline \noalign{\vspace{2pt}}
        Experienced a breach & 11 \\  
        \noalign{\vspace{2pt}} \hdashline \noalign{\vspace{2pt}}
        Sold data & 10 \\
        \noalign{\vspace{2pt}} \hdashline \noalign{\vspace{2pt}}
        Asked for student information & 6 \\  
        \noalign{\vspace{2pt}} \hdashline \noalign{\vspace{2pt}}
        Heard about issues second-hand & 5\\
        \noalign{\vspace{2pt}} \hdashline \noalign{\vspace{2pt}}
        Administrators warned about security issues & 2\\
        \bottomrule
    \end{tabular}
\end{table}

%\SummaryBox
%{Educators are mostly unaware of data deletion upon terminating use of \utech and, when they are aware, do not always use this feature.}

\subsection{Administrators' S\&P concerns with \utech} 
\label{sec:it_observed}

\smallskip
\noindent
\textbf{Privacy and security incidents.}
In the admin survey, we asked participants about security incidents that resulted from the use of shadow IT; three participants described such cases. 
A2 noted, \qoute{``We have had third-party integrations scrape user data without being vetted and without contractual controls in place.''} 
The data exposed included names, emails, and student IDs. As the admin put it, this was \qoute{``information commonly available to instructors,''} but the access by third-party tools without consent raised concerns.

Another issue was that faculty routinely auto-forward emails to personal emails. \qoute{``This became a major problem''}, as A9 noted \qoute{``None of our phishing controls or detection would detect anything after such emails had transited outside our managed environment, including phishing attack successes and account compromise.''} This shows how routine behavior inadvertently undermines institutional protections.

A17 mentioned a more serious incident involved a breach in the ticketing platform AudienceView. They shared \qoute{``the product was purchased by an affiliate and did not go through our procurement process. The affiliate org handled the breach notifications.''}

% Overall, these incidents reflect the real and ongoing security and privacy challenges admins face when unsanctioned apps are used without institutional oversight and proper review.

\smallskip
\noindent
\textbf{Increased risks because of unsanctioned apps.}
In addition to past incidents, administrators shared additional risks from \utech they anticipate (AQ.$8$). Two authors coded the $21$ responses and identified concrete risks (mismatches were resolved via discussions). Table~\ref{tab:risks_raised_by_admins} lists those risks; below, we discuss their overarching themes and contextualize them using participant quotes.

\begin{table}[h]
    % \footnotesize
    % \renewcommand{\arraystretch}{1.2}
    \centering
    \caption{Concerns (regarding \utech use) identified from (n=21) administrators' responses. Note that each response stated multiple concerns.}
    \begin{tabular}{p{10cm} p{1cm}}
        \midrule \midrule
    \textbf{Concerns} & \textbf{Count}  \\
    \midrule
    Data leakage & 13 \\
    \noalign{\vspace{2pt}} \hdashline \noalign{\vspace{2pt}}    
    Privacy invasion & 9\\
    \noalign{\vspace{2pt}} \hdashline \noalign{\vspace{2pt}}
    Increased exposure of institutional ecosystem to threats & 6\\
    \noalign{\vspace{2pt}} \hdashline \noalign{\vspace{2pt}}
    Increase regulatory or contractual liability & 5\\
    \noalign{\vspace{2pt}} \hdashline \noalign{\vspace{2pt}}
    Increased operational overhead and liability & 5\\
    \noalign{\vspace{2pt}} \hdashline \noalign{\vspace{2pt}}
    Lack of visibility & 4\\
    \noalign{\vspace{2pt}} \hdashline \noalign{\vspace{2pt}}
    Malware, backdoors, apps with stealthy behavior & 4\\
    \noalign{\vspace{2pt}} \hdashline \noalign{\vspace{2pt}}
    Lack of control over the system or data & 3\\
    \noalign{\vspace{2pt}} \hdashline \noalign{\vspace{2pt}}
    Lack of IT/vendor support and security patches & 3\\
    \noalign{\vspace{2pt}} \hdashline \noalign{\vspace{2pt}}
    (Rouge) AI tools use & 2\\
    \noalign{\vspace{2pt}} \hdashline \noalign{\vspace{2pt}}
    Intellectual property theft & 2\\
    \noalign{\vspace{2pt}} \hdashline \noalign{\vspace{2pt}}
    Data loss and corruption & 1\\
    \noalign{\vspace{2pt}} \hdashline \noalign{\vspace{2pt}}
    Performance degradation & 1\\
    \noalign{\vspace{2pt}} \hdashline \noalign{\vspace{2pt}}
    Loss of reputation & 1\\
    \noalign{\vspace{2pt}} \hdashline \noalign{\vspace{2pt}}
    Misplaced trust & 1\\
    \noalign{\vspace{2pt}} \hdashline \noalign{\vspace{2pt}}
    Silo effect (departments operating in isolation)  & 1\\
    \bottomrule
    \end{tabular}
    \label{tab:risks_raised_by_admins}
\end{table}

\smallskip
\noindent
\textbf{Threat to private data and increased liabilities.}
Administrators expressed the highest concerns about leaking sensitive data, which may happen in different stages, as A2 explained, \qoute{``Inadvertent reading, storing and displaying of information...especially where it is held by a third party with no legal agreement or controls in place... [if] the third party is breached, or [they] use such information for other purposes (selling data, marketing, use to improve their product, etc).''}
Several administrators stated that such access to institutional data is a clear violation of data protection laws and institutional policies. A19 emphasized, \qoute{``Disclosure of covered information to third parties not under contract...this can commonly meet the criteria for a FERPA breach exposing the institution to serious liabilities.''} Beyond federal regulations, educators' use of apps can lead to financial obligations for the institute:
A19 added, \qoute{``There can be contractual liabilities where now you've obligated the organization to acquire the license due to your use.'' }

\smallskip
\noindent
\textbf{Increased exposure to invisible threats.}
\label{invis_threats}
Administrators worried that \utech created \emph{new risks} because they can be faulty or deliberately malicious stealing personal data and intellectual property. A lack of visibility makes matters worse, as A2 highlighted \qoute{``No visibility into risk introduced... No information of [the] security posture of the technology.''} A11, added that these tools \qoute{``strip the controls [the institution] very intentionally design, build in and manage.''}

Such risks were heavily discussed in the context of third-party ``add-ons'' and AI-based plugins that can be easily ``added'' into other apps. A15 referenced \qoute{``insidious and rogue AI from third-party vendors in Zoom,''} and A16 highlighted the blind spots introduced by unvetted software: \qoute{``Not knowing what the application is doing and what backend communications it may establish.''} Similarly, A17 warned that \qoute{``The plug-in/software could be compromised to purposefully steal student data.''} From this inadvertent support from admins and educators, students may misplace their trust in these apps, thinking that they were vetted or managed by their school.

These risks were not unfounded as shadow IT is frequently being integrated into the institutional technology ecosystem (Q.$4d$). In our data, K-12 and HEI educators reported integrating $341$ applications into their accredited systems, and HEI educators integrated $116$ apps. These apps not only create blind spots but also introduce new vulnerabilities to the EdTech landscape by not being properly vetted~\cite{kelso2024trust}. Additionally, $338$ apps from K-12 and $140$ apps from HEIs required students to register using their institutional credentials (Q.$4e$), creating risks for being identified (and subsequently being surveilled or tracked).

%\SummaryBox
%{IT Administrators are concerned about introducing invisible S\&P vulnerabilities and non-compliance issues due to \utech{} use.}

\subsection{Risks from regulatory non-compliance}\label{sec:privacy:compliance}
 
In the US, FERPA~\cite{ferpaLaw} and HIPAA~\cite{hipaa} guide how educational and health records, respectively, can be used. But they do not typically cover digital tools from private entities unless they enter into contracts with education or health providers~\cite{kelso2024trust}. The developers, however, can proactively comply with these laws (like Google and Zoom). Educators are expected to be aware of such regulations and comply when handling students' personal data; violations could result in losing school funding, their job, or both. To investigate this in the context of using \utech (that can collect such data), we asked them if they believed the apps they used comply with FERPA or HIPAA (Q.$4p$ and Q.$4q$). We found that $107$ educator participants believed all apps they listed were FERPA-compliant, $237$ believed at least one was compliant, and $7$ believed none were. 
Notably, $71$ participants were unsure whether any of their apps were FERPA-compliant, and $44$ participants ($21$ of whom were K–12 educators) explicitly stated that at least one app was not FERPA-compliant. For HIPAA, $81$ participants believed all apps were compliant, $182$ believed at least one was, and $10$ believed none were. $51$ participants were unsure about HIPAA compliance, and $55$ participants (including $23$ K–12 educators) stated that at least one app was not HIPAA-compliant. Uncertainty about legal compliance, however, did not result in the discontinuation of using them.

% \SummaryBox{Educators overwhelmingly continue to use \utech even when they are aware that the \utech is not compliant with legal regulations.}

% Despite this uncertainty, only two participants reported discontinuing the use of all their apps. The remaining $373$ participants continued using at least one application, even when unsure of or explicitly questioning its compliance with applicable data protection laws.

Since applicability and compliance information are not always readily apparent, we manually reviewed the privacy policy and terms of service of the top $14$ apps used by a majority ($52\%$) of the surveyed educators. Among these, $8$ indicated that they comply with FERPA (Blooket, Canva, Kahoot!, EdPuzzle, IXL, Prodigy, Zip Grade, ABC Mouse). Five apps---Quizlet, Khan Academy, Quizizz, Seesaw, and Padlet--- stated to be FERPA-compliant only when licensed institutionally. Only Teachers Pay Teachers (TPT) seemed to not comply with FERPA. Three apps: TPT, IXL, and Padlet, explicitly mentioned \textit{surveilling} or \textit{tracking} users.  %(Blooket~\cite{Poly_blooket}, Canva~\cite{Poly_Canva}, Kahoot!~\cite{Poly_kahoot}, EdPuzzle~\cite{Poly_Ed}, IXL~\cite{Poly_IXL}, Prodigy~\cite{Poly_prodigy}, Zip Grade~\cite{Poly_zip}, ABC Mouse~\cite{Poly_abc}). Five apps---Quizlet~\cite{Poly_Quizlet}, Khan Academy~\cite{Poly_khan}, Quizizz~\cite{Poly_Quizizz}, Seesaw~\cite{Poly_seesaw}, and Padlet~\cite{Poly_padlet}--- stated to be FERPA-compliant only when licensed institutionally. Only Teachers Pay Teachers (TPT)~\cite{Poly_TPT} seemed to not comply with FERPA. Three apps: TPT, IXL, and Padlet, explicitly mentioned \textit{surveilling} or \textit{tracking} users. %Notably, all of these apps are focused on education services or have a separate ``student version''. 
COPPA most likely applies to each of these applications as students in K-12 were being registered for them; However, Khan Academy and Padlet stated that they only follow COPPA for verified student accounts. 

% These findings shed light on additional issues due to apps not being institutionally sanctioned, as that would require them to comply with privacy laws and may additionally restrict app data collection and use~\cite{kelso2024trust}.

%\SummaryBox{Use of \utech risks institutional regulatory non-compliance, in general. Interestingly, almost half of the most-used \utech applications \emph{do not} comply with FERPA and/or COPPA when used as \utech.}

\SummaryBox{\textbf{RQ3 Summary:} Shadow IT use risks regulatory non-compliance, and increases the possibility of invisible vulnerabilities into the institution. }

\section{Discussion}
In this section, we discuss the implications of our findings on various stakeholders and make recommendations based on our study's results to improve the status quo of utilizing \utech.

\subsection{Implications of our findings}

% \SR{I think we should highlight (make them more prominent) the implications of our findings. Currently, it is buried between our and past research findings.}

% \SR{We can break it into user groups, like, ``Implications for Educators'', ``Implications for Students''``Implications for Educational Institutes'', ``Implications for Regulators'', ``Implications for the Society'', etc.}

\smallskip
\noindent
\textbf{Implications for educators.} 
The continued use of \utech was commonplace, despite experiences of invasive behaviors, lack of trust on app developers, and availability of institutional alternatives. One key reason behind this was familiarity with the tools and an unwillingness to learn new ones (\S\ref{edtech_selection_criteria}). Such a prominent role of \textit{habituation} was found in other contexts as well (e.g., corporate employees~\cite{van2024department1}). However, in the corporate settings, typically a cold relationship exists between employees and admins who actively look to halt all aspects of \utech use when found~\cite{silic_shadow_2014, kopper2017perceptions}. In contrast, we found admins helping tool integration or even encouraging their adoption. In our case, it even led to the use of apps and services that were sued by the FTC for privacy violations~\cite{FTC-Edmodo}, and apps that were labeled as ``critically risky'' by external evaluators (e.g., Quizlet~\cite{AppMicro}). Educators are also overworked, often giving up personal activities to teach~\cite{Teach_Overwork}, which could by a key reason for shadow IT usage, as a way to lighten their load.
%which could be a key reason for high rates of shadow IT, as a way to lighten their load.
Another key reason might be a lack of knowledge of privacy and safety risks from app providers and developers~\cite{hielscherEmployeesWhoDont-20231}.
%Many participants lacked cybersecurity and privacy knowledge, such as an understanding of online tracking and data brokers, which are prerequisites to considering privacy risks as an app selection factor and judging apps based on how much risk they might pose.
% still tried to keep to good practices such as using passwords and regularly updating software; however, participants who had more knowledge were shown to be more vigilant, by detecting privacy-invasive behaviors and stopping the use of an app that demonstrated them. \textit{This implies that greater cybersecurity awareness leads to more proactive, privacy-protective behaviors.}

Despite lacking knowledge, Educators who value other people's privacy were more likely to consider security and privacy aspects before adopting a \utech (\S\ref{sec:personalfactors}). This suggests that raising privacy awareness among educators, particularly from the perspective of their role as \textit{loco parentis}, can discourage shadow IT adoption or encourage careful review (e.g., of expert evaluations~\cite{AppMicro}) to find better alternatives. Future studies could explore interventions to increase awareness of others' right to privacy.

\smallskip
\noindent
\textbf{Implications for students.}
Most of the apps and services identified were ``free''; however, they likely rely on advertisements for revenue generation. Personalized advertisement encourages (more) private data collection, and ads might expose students to inappropriate content or scams~\cite{shaoUnderstandingInAppAds-2018}. Indeed, our sample of educators reported many privacy invasive behaviors they observed (\S\ref{observed_risks}). Beyond immediate privacy and safety threats, we want to highlight some longer-term, perhaps more damaging impacts on students: normalizing surveillance and devaluing students' agency. Students, especially young ones, may become normalized to being surveilled, quantified, and compared to others, which can inhibit self-expression and encourage self-censoring behaviors, negatively impacting their development~\cite{achieving-privacy-in-education-2018, datafication-HE}. Prior studies reported students' discomfort even with the use of institutionalized technology~\cite{ multi-stakeholder-la, do-they-care, yangDiscoveringPrivacyHarms2024, student-perception-privacy, students-privacy-perception-proctoring}; imposing more technology use, especially against institutional policy, might send a signal that their agency and safety are not valued. Constant surveillance can create privacy fatigue~\cite{privacy_fatigue_2017} and lead to privacy resignation~\cite{resignation-app, corporate-digital-resignation}. Moreover, big corporations, which provide many critical services (including education platforms, job search and professional networking platforms, health portals, and insurance) and can be future employers, may obtain students' data through business agreements~\cite{zeideLearnerPrivacyMOOCs2018} or data marketplaces~\cite{student-data-broker, russell2018transparencystudentdata, student-data-sold}, opening doors for exploitation and discrimination~\cite{koizumiUnbearableDarknessBeing2026}. Lastly, machine learning models can be used to leak (by prediction) private information; e.g., interactions with digital tools have been shown to reveal demographics (e.g., gender and age~\cite{edtech-pets22}) and personality traits~\cite{peltonenPredictPersonalityMobileApp2020}, raising severe privacy, safety, and ethical concerns~\cite{kyritsi2019pursuit, edm_and_privacy}.

\smallskip
\noindent
\textbf{Implications for institutions.}
We identified several potential ways in which \utech use can negatively impact the privacy, security, and compliance postures of educational institutions (\S\ref{sec:it_observed}).
Here, we discuss multiple main drivers of continued \utech use despite the associated risks. Past research highlighted the influence of workplace environment and peer behavior on encouraging learning and complying with cybersecurity measures~\cite{liInvestigatingImpactCybersecurity-2019}. 
Results from our study portray a permissive environment at educational institutes. With an increasing workload, complying with or even becoming aware of privacy requirements sets a high bar. 
Administrators apparently lacked any real means or resources to enforce those policies, while many encouraged \utech adoption. This creates a negative network effect as people prefer platforms that are popular among social and professional ties to privacy-friendly alternatives~\cite{zhangPrivacyVsConvenience-2024}, continuously worsening the situation. This is critical because even if an app developer is at fault for misusing students' data, typically, it is the institutions that will be held liable by regulators, and they also stand to suffer from reputational damage.

\smallskip
\noindent
\textbf{Broader societal implications.}
The broader tech-ubiquitous environment of the education sector~\cite{datafication-HE, terasPostCovid19EducationEducation-2020, heads-in-the-clouds} may drive \utech adoption even by people who are aware of the risks. This is because being surrounded by technology can create a feeling of powerlessness over personal information~\cite{hargittai2016can1}; people may become ``privacy cynics'' where they abandon efforts and lose interest in privacy because the data will be out anyway~\cite{shalawadiDrConvenienceLove-PrivacyApathy2024}. Indeed, such coping mechanisms are not only common in the general population~\cite{ shalawadiDrConvenienceLove-PrivacyApathy2024, hargittai2016can1}, but also among data stewards in organizations~\cite{popescuRolebasedPrivacyCynicism-2024}. This may explain why educators, who were aware of the risks, as well as administrators, were accepting of \utech---if there is no escape, there is no point in rejecting the convenience these tools can provide. Moreover, when 96\% of school-approved apps send data to advertisers~\cite{apps-sell-data}, educators (and students) may feel demotivated in taking privacy-protective steps.

\subsection{Recommendations.}

\smallskip
\noindent
\textbf{Privacy-orientation and policy enforcement.} 
Valuing students' privacy was a key differentiator in educators \utech selection behaviors, but awareness about how privacy can be invaded was infrequent. Thus, besides security threats from external actors, cybersecurity awareness campaigns, training materials, and tutorials should also prioritize teaching privacy-related concepts and raise awareness about the additional risks that \utech creates. Often, privacy and security mechanisms are perceived as barriers or disturbances to actual tasks~\cite{cram2023waste, sasseRebootingITSecurity-2023};
awareness or training campaigns can be contextualized to highlight how a privacy-respecting environment may foster freedom of expression and critical thinking among students, which are among the key educational goals. A lack of awareness of institutional policy was too common, and their disregard was also notable; while in many cases it was because of convenience, in other cases, it was out of necessity. Thus, with stricter policy enforcement, it is also essential to work with educators to find an acceptable solution, such as providing a temporary, isolated environment to use a service, or helping them move to a more privacy-friendly alternative without disrupting their workflow~\cite{zhangPrivacyVsConvenience-2024}. 
% (recommendation for pets developer).

% Only raising awareness is insufficient for actions if people do not have resources/lack capabilities to follow through~\cite{badaCyberSecurityAwareness-2019} or the process 

% Seminars and tutorial sessions to teach how sensitive data collection, even by service providers, risks privacy and safety can make a big difference in \utech adoption and use. Commercially available training packages exist, but these generic and checkbox solution oriented nature may not effectively serve schools with diverse needs and structures~\cite{hielscherEmployeesWhoDont-2023}. While customizing training packages for each institute may also not be feasible, 

\smallskip
\noindent
\textbf{Cultivating a privacy-respecting environment.} To sustain privacy-protective behaviors, schools must create an environment where such behaviors are rewarded. A culture of open discussion about privacy and related concepts within organizations drives behavior change~\cite{chaudharyDrivingBehaviourChange-2024}. This is particularly important in this context as educators source \utech from peers and past research has shown that social and professional ties shape attitude and behaviors~\cite{zhangPrivacyVsConvenience-2024}. We recommend that privacy-conscious educators be assigned roles like ``privacy champions''~\cite{becker2017findingsecuritychampion}, who will transfer that knowledge to peers in casual settings and be rewarded for such efforts. Such local efforts have multiple benefits. Their knowledge is tuned to the local environment and more relevant to their peers. Unlike formal training by external experts, local learning sessions can be frequent, which can help replace past habits with new, privacy-protective behaviors~\cite{hielscher_taking_2022} and embed them into the workflow~\cite{sasseRebootingITSecurity-2023}. Local privacy champions can share their knowledge and recommendations in a casual setting that is more likely to be valued than formal warnings (which, our results suggests, get regularly ignored). Finally, these local efforts will, over time, have the potential to diminish the negative network effect~\cite{zhangPrivacyVsConvenience-2024} and help establish a privacy-respecting environment.  %local technology

\smallskip
\noindent
\textbf{Encourage critical evaluation of apps.} Equipped with privacy literacy and risk awareness, educators should be encouraged to evaluate apps before use. Since peer network plays an important role in \utech use, administrators and peers, knowledgeable in this matter, can point to obvious indicators, such as asking for too many permissions or lacking a privacy policy. They can also identify and popularize external, more technically comprehensive assessments that are available online from authoritative organizations (e.g.,~\cite{AppMicro, scaryapps, ikeepsafe}).

\section{Conclusion}
We unveil the ubiquitous use of shadow IT across K-12 and HEI institutions, not born out of malicious intent, but rather of necessity. Other factors, such as a lax environment, non-existent or inconsistent guidelines, and resource constraints, are also to blame. Crucially, collaborating with educators---not simply enforcing compliance---will be essential in building a safer and more sustainable digital learning ecosystem. We conclude by stressing the urgent need for more research to identify effective ways to raise privacy awareness and encourage thoughtful technology integration through privacy campaigns, especially from the perspective of how educators' decisions not only affect students' privacy, but also shape their attitude towards privacy. We suggest that, to succeed, these campaigns must connect the importance of privacy to values embedded in the education context, such as self-expression and freedom of thought.

% of the risks associated with unsanctioned technologies and foster a privacy-respecting environment that incentivizes the use of approved tools. 

%Educational institutions must invest in creating educator-friendly,
%while secure alternatives and provide clear pathways for vetting and approving education tools. Moreover, fostering
%a culture of shared responsibility—where both educators and administrators collaborate to balance innovation with
%compliance—will be essential for addressing these challenges effectively

\section*{Acknowledgments}
This was was supported in part by the National Science Foundation under Grants 2544663 and 2350036.

%Bibliography
\bibliographystyle{unsrt}

\end{document}